\documentclass[aps,twocolumn,groupedaddress,floats,showpacs]{revtex4}
\usepackage{graphicx}
\usepackage{bm}
\begin{document}

\title{Supersolid phase of hardcore bosons on triangular lattice.}
\author{Massimo Boninsegni$^{1}$ and Nikolay Prokof'ev$^{2,3}$}
\affiliation{ ${^1}$Department of Physics, University of Alberta,
Edmonton, Alberta T6G 2J1\\
${^2}$Department of Physics, University of
Massachusetts, Amherst, MA 01003 \\
${^3}$Russian Research Center ``Kurchatov Institute'', 123182
Moscow }

\begin{abstract}
We study properties of the supersolid phase observed for hardcore
bosons on the triangular lattice near half-integer filling factor,
and the phase diagram of the system at finite temperature. We find
that the solid order is always of the $(2m,-m',-m')$ with $m$
changing discontinuously from positive to negative values at
half-filling, in contrast with phases observed for Ising spins in
transverse magnetic field. At finite temperature we find two
intersecting second-order transition lines, one in the $3$-state
Potts universality class and the other of the Kosterlitz-Thouless
type.

\end{abstract}

\maketitle

Since the supersolid state of matter was introduced to physics
nearly half a century ago and its theoretical feasibility was
demonstrated,\cite{Penrose}
there was a long history of
experimental attempts to find it in Nature (mostly in $^4$He, see,
e.g., Ref. \onlinecite{Meisel}) along with numerical simulations and
theoretical predictions for models of interacting lattice bosons.
Recent years have seen a renewed interest in this topic. On the one
hand, lattice bosons are no longer in the realm of idealized
models and can be now studied in controlled experiments with
ultra-cold atoms in optical potentials \cite{Greiner}. On the
other hand, the non-classical moment of inertia observed for solid $^4$He
samples in the torsional oscillator experiments by Kim and Chan
\cite{Kim} remains largely a mystery.

Hardcore bosons on triangular lattice with nearest-neighbor
repulsion $V>0$ and hopping $t>0$ represent one of the simplest
(and thus most promising from the experimental point of view)
models displaying a supersolid phase in an extended region of the
phase diagram. The model Hamiltonian is given by:
\begin{equation}
H=-t\sum_{\langle ij\rangle}(\hat b_i^{\dagger}\hat b_j{\;} + h.c) + V\sum_{\langle ij\rangle} \hat n_i\hat n_j
- \mu \sum_{i}\hat n_i\;. \label{H}
\end{equation}
Here $\hat b_i^{\dagger}$ is the bosonic creation operator,
$\hat n_i=\hat b_i^{\dagger}\hat b_i{\;}$, and $\mu$ is the chemical potential.
A triangular lattice of $N=L\times L$ sites, with periodic boundary conditions is assumed.
The alternative formulation of (\ref{H}) in terms of quantum spin-1/2 variables
$\hat s_i$, namely
\begin{equation}
H=-2t\sum_{
\langle ij\rangle}(\hat s_i^{x}\hat s_j^{x} +\hat s_i^{y}\hat s_j^{y} ) + V\sum_{\langle ij\rangle}
\hat s_i^{z}\hat s_j^{z} - (\mu-3V) \sum_{i}\hat s_i^{z}\ \label{Hs}
\end{equation}
provides a useful mapping to the XXZ-magnet. The superfluid state
of Eq.~(\ref{H}) for $t>>V$ corresponds to the XY-ferromagnetic
state of Eq.~(\ref{Hs}), while the solid state of bosons is
equivalent to magnetic order in the $\hat{z}$-direction. At
half-integer filling factor, $n(\mu=3V) =1/2$, the model has an
exact particle-hole symmetry.

\begin{figure}[tbp]
\centerline{\includegraphics[bb=30 30 500 750, angle=-90, width=1.\columnwidth]{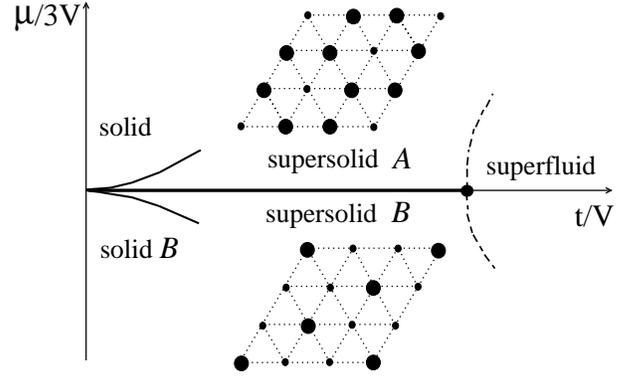}}
\caption{Schematic phase diagram of Eq.~(\ref{H}) near half-integer
filling factor.}
\label{fig1}
\end{figure}

A robust confirmation of early mean-field predictions of a
supersolid phase in the ground state of (\ref{H}), \cite{Murthy}
was obtained by means of Green function Monte Carlo (GFMC)
simulations.\cite{Boninsegni} The supersolid phases identified in
that study for densities away from half-filling (i.e., for $\mu/V
>3$ and $\mu/V<3$), can be viewed as solids, with filling factors
$\nu=2/3$ and $\nu=1/3$, doped with holes  and particles
respectively. In what follows, we denote them as supersolids
${\cal A}$  and ${\cal B}$. Density correlations in ${\cal A}$ and
${\cal B}$ have $\sqrt{3}\times \sqrt{3}$ ordering with the wave
vector ${\bf Q}=(4\pi /3 ,0)$. In ${\cal A}$ and ${\cal B}$ the
average occupation numbers on three consecutive sites along any of
the principal axes follow the sequence $(-2m,m',m')$ and
$(2m,-m',-m')$ respectively (it is conventional to count densities
from $1/2$ to make connection with the magnetization in the spin
language, $m_i=n_i-1/2$), see Fig.~\ref{fig1}.

The model (\ref{H}) has been investigated in a series of recent
papers, making use of advanced numerical
techniques.\cite{Kedar,Melko,Wessel} The proposed zero-temperature
phase diagram is similar to that of Ref. \cite{Boninsegni},  with
the notable addition of a quantum superfluid-supersolid phase
transition at $n$=1/2 and $t/V\approx 0.115$ and  the stable
supersolid state persisting for smaller values of $t/V$. In Ref.
\cite{Boninsegni} the system was thought to remain a disordered
superfluid for arbitrary $t$/$V$. The discrepancy can be
attributed to known limitations of the GFMC method. \cite{Note}

Based on field-theoretic, exact diagonalization,  and other
arguments, Ref.~\cite{Melko} hints at the possibility of the
$(m,0,-m)$ density order in the ground state at $n=1/2$ (state
${\cal C}$). These considerations involved, in particular, an
analogy between the properties of Eq.~(\ref{Hs}), and those of the
Ising antiferromagnet on the triangular lattice, in the presence
of a transverse magnetic field \cite{Isakov}. If true, there
should exist quantum ${\cal A}-{\cal C}$ and ${\cal C}-{\cal B}$
phase transitions away from half-filling and three
finite-temperature transitions of the Kosterlitz-Thouless (KT)
type.
Though Ref.~\cite{Kedar} finds that the ground state is of the
${\cal A}$ or ${\cal B}$ type, it makes
similar predictions for the finite temperature phase diagram at
$n=1/2$ which follow from the assumption that spontaneous symmetry
breaking between ${\cal A}$, ${\cal B}$, and their lattice
translations is described by the six-clock model \cite{Jose}.

In what follows, we provide strong evidence that the supersolid
state at half-filling is always of either the ${\cal A}$ or ${\cal
B}$ type. Our data suggest that there is a discontinuous
transition from ${\cal A}$ to ${\cal B}$ at $\mu=3V$ similar to
the I-order phase transition (driven by the large energy of the
${\cal A-B}$ domain walls). What makes it special is the exact
particle-hole symmetry; structure factor, superfluid density, and
energy remain continuous functions of $\mu$ through the transition
line. For the supersolid ${\cal A}$ (or ${\cal B}$) with the
three-fold degenerate ground state, one expects to see the
normal-superfluid KT and the solid-liquid $3$-state Potts
transitions, as temperature is increased. Moreover, the KT and
Potts transitions are independent of each other and for $n \ne
1/2$ intersect on the phase diagram. The failure of the mean-field
description and analogies with the transverse-field Ising model to
predict the supersolid structure at $n=1/2$ can be traced back to
the U(1)-symmetry of Eqs.~(\ref{H}) and (\ref{Hs}), as noticed in
\cite{Kedar}. For example, the $(1,0,-1)$ state can {\it not} be
the true groundstate at finite $t$ in the limit of $t/V \to 0$
simply because it does not respect the particle conservation law.

\begin{figure}[tbp]
\centerline{\includegraphics[bb=30 30 570 790, angle=-90, width=1.\columnwidth]{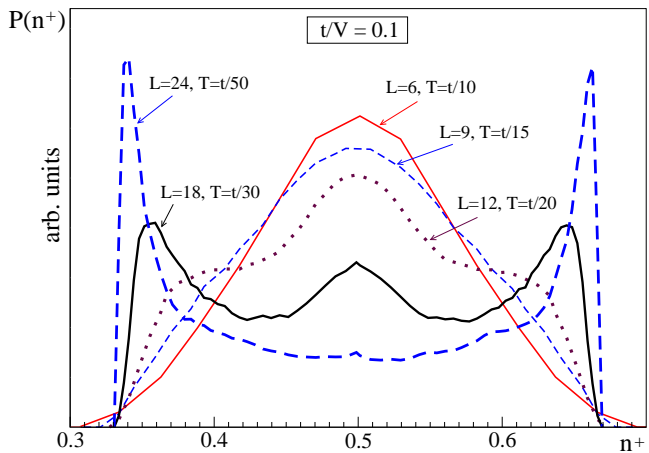}}
\caption{ (Color online). Probability distributions $P(n^{+}) $ for different system sizes
and temperatures at $\mu/V=3$ and $t/V=0.1$.}
\label{fig2}
\end{figure}

\begin{figure}[tbp]
\centerline{\includegraphics[bb=30 30 570 790, angle=-90, width=1.\columnwidth]{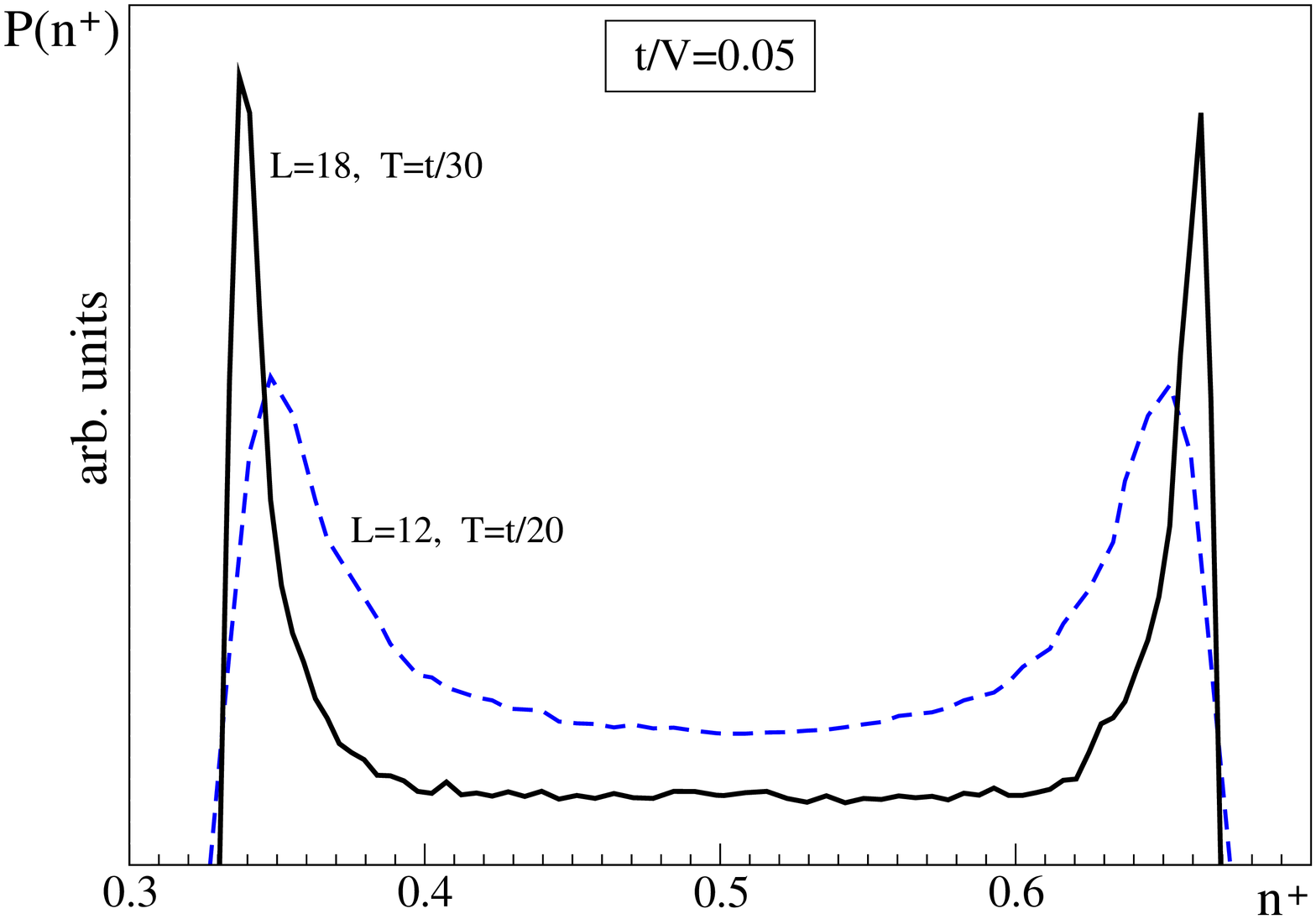}}
\caption{ (Color online). Probability distributions $P(n^+) $ for
different system sizes
and temperatures at $\mu/V=3$ and $t/V=0.05$. }
\label{fig3}
\end{figure}

We use the worm-algorithm Monte Carlo scheme in the lattice path-integral
representation \cite{worm} to simulate Eq.~(\ref{H}).
Since the structure factor
\begin{equation}
S({\bf Q})= \biggl \langle \bigg| \sum_{k=1}^{N} \hat n_k\
e^{i{\bf Qr}_k} \bigg|^2 \biggr \rangle/N^2
\end{equation}
does not distinguish between supersolids ${\cal A,\ B,\ C}$,
we adopt the following strategy: for
each system configuration, we compute the distribution of
time-averaged occupation numbers,
$\bar{n}_k=\beta^{-1} \int_0^{\beta } d\tau\ \hat n_k(\tau )$,
and use it to determine the fraction of sites with $\bar{n}_k>1/2$
\begin{equation}
n^{+}= N^{-1} \sum_{k=1}^{N} \theta ( \bar{n}_k -1/2 )\;,
\label{np}
\end{equation}
where $\theta(x)$ is the Heviside function.  ${\cal A,\ B,\ C}$
density structures correspond to $n^{+}_{A}=2/3$, $n^{+}_{C}=0$,
and $n^{+}_{B}=1/3$. Finite systems are characterized by broad
probability distributions $P(n^+)$, and the formation of different
solid orders can be seen as the development of sharp peaks, as the
thermodynamic limit is approached.

In Fig.~\ref{fig2} we show the evolution of the $P(n^{+})$
distribution for the half-filled system at $V/t=10$, i.e., close
to the superfluid-supersolid transition point,
estimated\cite{Kedar,Melko,Wessel} at $V/t \approx$ 8.5.  The
distribution is peaked at $n^{+}=0$ in the smallest system
considered ($L$=6), but, as the system size is increased, the
weight is shifted toward the wings of the distribution. For
$L$=18, there are already three peaks with comparable height.
Finally, in the $L$=24 system we observe only two peaks
corresponding to the supersolid phases ${\cal A}$ and  ${\cal B}$.
Though the probability density between the peaks is still
measurable, the dynamics of the algorithm becomes very slow; it
typically takes millions of Monte Carlo sweeps, in order for the
system to make a transition from the  ${\cal A}$ to the ${\cal B}$
structure and vice versa. We have explicitly verified that
configurations with $n^{+} \approx 2/3$ and $n^{+} \approx 1/3$
have density orders depicted as in Fig.~\ref{fig1}, with a large
contrast in density between sublattices. We have also checked that
the $V/t=10$, $L$=48, $T$ = $t/$50 system spontaneously develops
either ${\cal A}$ or ${\cal B}$ order, starting from the initial
configuration corresponding to the superfluid phase at $V/t$=5.

\begin{figure}[tbp]
\centerline{\includegraphics[bb=30 30 570 790, angle=-90, width=1.\columnwidth]{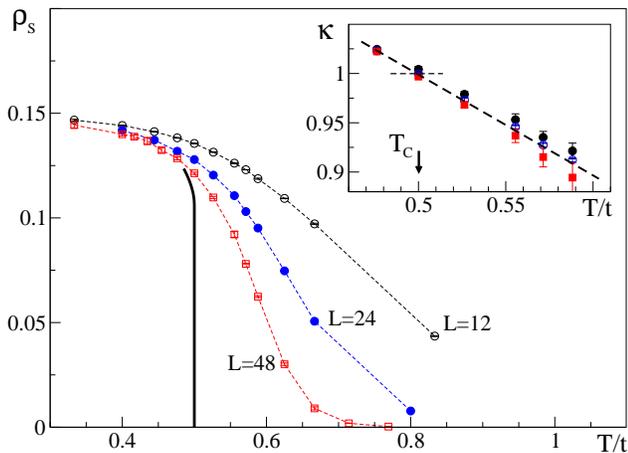}}
\caption{(Color online). Superfluid density in the vicinity of the KT transition for $t/V=0.1$
and $\mu/V=2.74$. The solid line is the thermodynamic curve calculated
using Eq.~(\ref{R}) with $\kappa (T)$ deduced from the plot shown in the inset.
Inset: solutions of the Eq.~(\ref{RG}) for different pairs of system sizes:
$L_2=24$, $L_1=12$---filled circles, $L_2=48$, $L_1=12$---open circles
$L_2=48$, $L_1=24$---filled squares. The dashed line is the linear fit
$\kappa=1+1.03(T_c-T)/t$ with $T_c/t=0.50$.}
\label{fig4}
\end{figure}

\begin{figure}[tbp]
\centerline{\includegraphics[bb=30 30 570 790, angle=-90, width=1.\columnwidth]{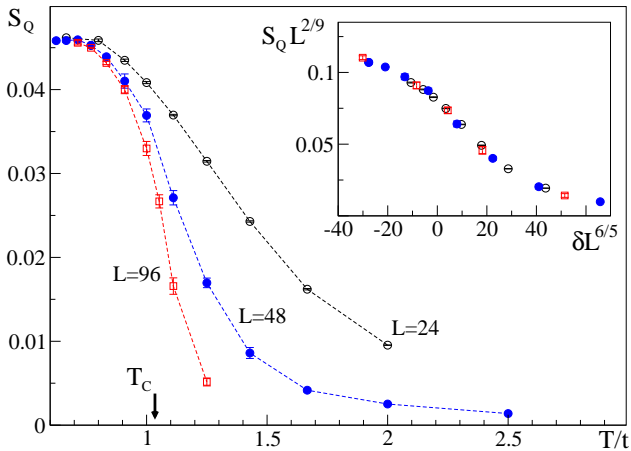}}
\caption{(Color online).  Structure factor in the vicinity of the 3-state Potts transition for $t/V=0.1$
and $\mu/V=2.74$.
Inset:  data collapse using exact critical exponents for the 3-state Potts model
\cite{Alexander}   and  $\delta =(T-T_c)/t$ with $T_c/t=1.035$.}
\label{fig5}
\end{figure}

In Fig.~\ref{fig3} we show what happens at larger values of $V/t$.
Now, the central peak is already absent in relatively small $L$=12 and $L$=18 systems.
We thus conclude that the nature of the supersolid state at half-integer filling factor
is determined by the ${\cal A}$ and ${\cal B}$  structures,
for {\it all} values of $t/V$ for which a supersolid phase exists.

If spontaneous symmetry breaking of the ground state degeneracies
is described by the six-clock model \cite{Jose}, one should
observe three finite-temperature transitions for systems near
half-filling, and a solid phase with algebraic correlations
``sandwiched" between the solid and normal liquid phases. This
prediction was made in Ref.~\cite{Kedar} for $n=1/2$. Since the
ground state was found here to be only of the ${\cal A}$ or ${\cal
B}$ type, and we do not see why domain wall energies between
translated ${\cal A}$ states are the same as between ${\cal A}$
and  ${\cal B}$ states (in fact, the Landau theory prediction
\cite{Melko,Kedar,Burkov} is that ${\cal A}$ and ${\cal B}$ states
phase separate and have different average densities even at
$\mu=3V$), the finite temperature phase diagram should instead
feature the normal-superfluid KT and the liquid-solid 3-state
Potts (for $n \ne 1/2$) transitions breaking U(1) and translation
symmetry respectively. At $n=1/2$ we expect only one liquid-solid
transition. An interesting question is whether transition lines
simply intersect, or there are bicritical and tricritical points
and I-order lines as observed for the similar model on the square
lattice \cite{square}. We performed simulations for two
representative cases, one for constant chemical potential
$\mu/V=2.74$ (or density $n\approx 0.44$), and the other for
constant $t/V=0.1$.

In Fig.~\ref{fig4} we show typical data for the KT transition between the solid
and supersolid phases. The transition is smeared by logarithmic finite-size effects,
but the critical  temperature can be still determined with good accuracy
by utilizing the well known renormalization flow and the universal jump of the
superfluid density, $\rho_s$, at $T_c$. The data analysis is as follows:\cite{weak2D}
we define  $R=\pi \rho_s /2mT$ (where $m=1/3t$ is the effective mass for the
triangular lattice) and study the finite-size scaling of the data using KT renormalization
group equations in the integral form
\begin{equation}
4 \ln (L_2/L_1) = \int_{R_2}^{R_1}  {dt \over t^2(\ln (t) - \kappa )+t }    \;.
\label{RG}
\end{equation}
The microscopic (system size independent) parameter $\kappa$ is an
analytic function of temperature, and the critical point corresponds to
$R=1$  at $\kappa=1$. For $T<T_c$, the thermodynamic curve is defined by the
equation
\begin{equation}
1/R+{\rm ln} \ R=\kappa (T) \;,
\label{R}
\end{equation}
with $\kappa = 1 + \kappa ' (T_c-T)$. We use different pairs of system sizes
in Eq.~(\ref{RG}) to determine the $\kappa (T)$ curve, and obtain the location
of the critical point from  $\kappa (T_c)=1$. The results are shown in the inset of
Fig.~\ref{fig4}.   Data collapse and smooth analytic behavior of $\kappa (T)$ proves that
the transition is indeed of the KT type. We used the same protocol and system sizes
to determine other critical points.

In Fig.~\ref{fig5}, we present our data for the transition into the state with the long-range
density order. For the three-fold degenerate ${\cal B}$ structure this transition
is expected to be in the 3-state Potts universality class. The critical exponents are
known exactly \cite{Alexander}: $\nu =5/6$, and $\beta = 1/9$. We thus perform the
data collapse using $L^{2\beta}S_{\bf Q} = f(\delta L^{1/\nu})$ where $\delta =(T-T_c)/t$ and
$T_c$ is  the only fitting parameter. The result is shown in the inset of Fig.~\ref{fig5}.
This  confirms the above-mentioned expectation, and establishes that there is only
one transition to the solid phase (there are no visible finite-size effects below $T_c$).

\begin{figure}[tbp]
\centerline{\includegraphics[bb=30 30 570 790, angle=-90, width=1.\columnwidth]{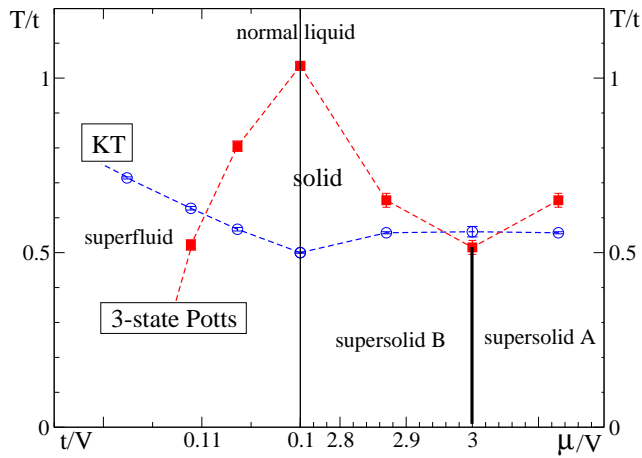}}
\caption{(Color online). Finite temperature phase diagram for two representative cuts:
the left panel is for fixed $\mu/V=2.74$; the right panel is for fixed $t/V=0.1$.
The solid line indicates a degenerate I-order transition line between supersolids
${\cal B} $ and ${\cal A} $. }
\label{fig6}
\end{figure}

Finally, we compute the phase diagram in the $(T/t,t/V)$ (at
constant $\mu/V=2.74$) and $(T/t,\mu/V)$ (at constant $t/V=0.1$)
planes and observe that KT and Potts transition lines  form a
simple cross for $n \ne 1/2$, i.e.,  the corresponding order
parameter fields are not strongly interacting, see
Fig.~\ref{fig6}. The transition temperature to the superfluid and
supersolid states in this part of the phase diagram is determined
by the hopping amplitude.  Within the statistical uncertainties of our
calculation, KT and Potts transition temperatures cannot be distinguished
at $\mu/V$=3.

We did not see evidence for the algebraic solid state at $\mu=3V$.
The finite-size scaling for the supersolid-solid transition at
$\mu=3V$ is consistent with the $3$-state Potts universality, though
the data collapse is not as impressive as in Fig.~\ref{fig4}
(the other alternative is the KT transition).

It is instructive to understand why the $(m,0,-m)$ phase for the
Hamiltonian (\ref{H}) is not an obvious groundstate. At the
mean-field level, ${\cal C}$ has a better energy than ${\cal A}$
or ${\cal B}$. For the transverse-field Ising model \cite{Isakov}
the $(1,0,-1)$ spin arrangement is obtained by orienting the
middle spin along the magnetic field direction, i.e. putting it in
the equal-amplitude superposition of up- and down-states. In
bosonic language, it corresponds to the superposition of states
with one or zero particles on a given site.  Such a state can not
be reconciled with the Hamiltonian (\ref{H}) which conserves the
particle number. Any non-integer average occupation number
necessarily involves hopping transitions to the nearest neighbor
sites. In the $(1,0,-1)$ structure the middle site is completely
surrounded by the fully occupied or empty sites and thus can not
be the ground state of the system even in the limit of $t/V \to
\infty$. The problem appears to be inherently quantum with no
obvious solution at the mean-field level.

We are grateful M. Troyer, A. Paramekanti, K. Damle, and A. Burkov
for stimulating discussions. The research was supported by the
National Science Foundation under Grant No. PHY-0426881, by the
Petroleum Research Fund of the American Chemical Society under
research grant 36658-AC5, and by the Natural Science and
Engineering Research Council of Canada under research grant
G121210893.

\end{document}